\documentclass[
    ,final            
  ]
  {aipproc}

\layoutstyle{6x9}

\begin{document}


\newcommand{\alps}{\ensuremath{\alpha_s}}
\newcommand{\qbar}{\bar{q}}
\newcommand{\beq}{\begin{equation}}
\newcommand{\eeq}{\end{equation}}
\newcommand{\beqa}{\begin{eqnarray}}
\newcommand{\eeqa}{\end{eqnarray}}
\newcommand{\gs}{g_{\pi NN}}
\newcommand{\gw}{f_\pi}
\newcommand{\mq}{m_Q}
\newcommand{\mn}{m_N}
\newcommand{\bb}{\langle}
\newcommand{\kb}{\rangle}
\newcommand{\st}{\ensuremath{\sqrt{\sigma}}}
\newcommand{\rvec}{\mathbf{r}}
\newcommand{\bvec}[1]{\ensuremath{\mathbf{#1}}}
\newcommand{\bra}[1]{\ensuremath{\bb#1|}}
\newcommand{\ket}[1]{\ensuremath{|#1\kb}}
\newcommand{\gft}{\ensuremath{\gamma_{FT}}}
\newcommand{\gfv}{\ensuremath{\gamma_5}}
\newcommand{\bfalp}{\mbox{\boldmath{$\alpha$}}}
\newcommand{\bfgam}{\mbox{\boldmath{$\gamma$}}}
\newcommand{\bfnab}{\mbox{\boldmath{$\nabla$}}}
\newcommand{\bflambda}{\ensuremath{\bm{\lambda}}}
\newcommand{\bfmu}{\ensuremath{\bm{\mu}}}
\newcommand{\bfphi}{\mbox{\boldmath{$\phi$}}}
\newcommand{\bfpi}{\mbox{\boldmath{$\pi$}}}
\newcommand{\bfsig}{\mbox{\boldmath{$\sigma$}}}
\newcommand{\bfrho}{\ensuremath{\bm{\rho}}}
\newcommand{\bftau}{\mbox{\boldmath{$\tau$}}}
\newcommand{\bfth}{\mbox{\boldmath{$\theta$}}}
\newcommand{\bfchi}{\ensuremath{\bm{\chi}}}
\newcommand{\bfR}{\ensuremath{\bvec{R}}}
\newcommand{\Rcm}{\ensuremath{\bvec{R}_{CM}}}
\newcommand{\spup}{\uparrow}
\newcommand{\spd}{\downarrow}
\newcommand{\hbarom}{\frac{\hbar^2}{m_Q}}
\newcommand{\half}{\ensuremath{\frac{1}{2}}}
\newcommand{\thalf}{\ensuremath{\frac{3}{2}}}
\newcommand{\shalf}{{\scriptstyle\frac{1}{2}}}
\newcommand{\sthalf}{{\scriptstyle\frac{3}{2}}}
\newcommand{\vij}{\ensuremath{\hat{v}_{ij}}}
\newcommand{\vnn}{\ensuremath{\hat{v}_{NN}}}
\newcommand{\argonne}{\ensuremath{v_{18}}}
\newcommand{\lqcd}{\ensuremath{\mathcal{L}_{QCD}}}
\newcommand{\lgf}{\ensuremath{\mathcal{L}_g}}
\newcommand{\lqm}{\ensuremath{\mathcal{L}_q}}
\newcommand{\lqg}{\ensuremath{\mathcal{L}_{qg}}}
\newcommand{\nn}{\ensuremath{N\!N}}
\newcommand{\hpnd}{\ensuremath{H_{\pi N\Delta}}}
\newcommand{\hpqq}{\ensuremath{H_{\pi qq}}}
\newcommand{\fpnn}{\ensuremath{f_{\pi NN}}}
\newcommand{\fpnd}{\ensuremath{f_{\pi N\Delta}}}
\newcommand{\fpqq}{\ensuremath{f_{\pi qq}}}
\newcommand{\ylm}{\ensuremath{Y_\ell^m}}
\newcommand{\ylmc}{\ensuremath{Y_\ell^{m*}}}
\newcommand{\qbh}{\hat{\bvec{q}}}
\newcommand{\xbh}{\hat{\bvec{X}}}
\newcommand{\rbh}{\hat{\bvec{r}}}
\newcommand{\zbh}{\hat{\bvec{z}}}
\newcommand{\dt}{\Delta\tau}
\newcommand{\kmag}{|\bvec{k}|}
\newcommand{\pmag}{|\bvec{p}|}
\newcommand{\qmag}{|\bvec{q}|}
\newcommand{\oas}{\ensuremath{\mathcal{O}(\alpha_s)}}
\newcommand{\vtxb}{\ensuremath{\Lambda_\mu(p',p)}}
\newcommand{\vtxp}{\ensuremath{\Lambda^\mu(p',p)}}
\newcommand{\pwqp}{e^{i\bvec{q}\cdot\bvec{r}}}
\newcommand{\pwqm}{e^{-i\bvec{q}\cdot\bvec{r}}}
\newcommand{\gsa}[1]{\ensuremath{\bb#1\kb_0}}
\newcommand{\oer}[1]{\mathcal{O}\left(\frac{1}{\qmag^{#1}}\right)}
\newcommand{\nub}[1]{\overline{\nu^{#1}}}
\newcommand{\balph}{\mbox{\boldmath{$\alpha$}}}
\newcommand{\bgam}{\mbox{\boldmath{$\gamma$}}}
\newcommand{\epf}{E_\bvec{p}}
\newcommand{\epfp}{E_{\bvec{p}'}}
\newcommand{\eka}{E_{\alpha\kappa}}
\newcommand{\ekaq}{(E_{\alpha\kappa})^2}
\newcommand{\ekap}{E_{\alpha'\kappa}}
\newcommand{\ekpa}{E+{\alpha\kappa_+}}
\newcommand{\ekma}{E_{\alpha\kappa_-}}
\newcommand{\ekp}{E_{\kappa_+}}
\newcommand{\ekm}{E_{\kappa_-}}
\newcommand{\ekpap}{E_{\alpha'\kappa_+}}
\newcommand{\ekmap}{E_{\alpha'\kappa_-}}
\newcommand{\yjm}[1]{\mathcal{Y}_{jm}^{#1}}
\newcommand{\ysa}[3]{\mathcal{Y}_{#1,#2}^{#3}}
\newcommand{\ysc}{\tilde{y}}
\newcommand{\enm}{\varepsilon_{NM}}
\newcommand{\Scg}[6]
	{\ensuremath{S^{#1}_{#4}\:\vphantom{S}^{#2}_{#5}
 	 \:\vphantom{S}^{#3}_{#6}\,}}
\newcommand{\Kmat}[6]
	{\ensuremath{K\left[\begin{array}{ccc} 
	#1 & #2 & #3 \\ #4 & #5 & #6\end{array}\right]}}
\newcommand{\irt}{\ensuremath{\frac{1}{\sqrt{2}}}}
\newcommand{\irth}{\ensuremath{\frac{1}{\sqrt{3}}}}
\newcommand{\irs}{\ensuremath{\frac{1}{\sqrt{6}}}}
\newcommand{\rtoth}{\ensuremath{\sqrt{\frac{2}{3}}}}
\newcommand{\Tg}{\ensuremath{\mathsf{T}}}
\newcommand{\irrep}[1]{\ensuremath{\mathbf{#1}}}
\newcommand{\cirrep}[1]{\ensuremath{\overline{\mathbf{#1}}}}
\newcommand{\Fij}{\ensuremath{\hat{F}_{ij}}}
\newcommand{\Fqij}{\ensuremath{\hat{F}^{(qq)}_{ij}}}
\newcommand{\Fsij}{\ensuremath{\hat{F}^{(qs)}_{ij}}}
\newcommand{\Opij}{\mathcal{O}^p_{ij}}
\newcommand{\titj}{\bftau_i\cdot\bftau_j}
\newcommand{\sisj}{\bfsig_i\cdot\bfsig_j}
\newcommand{\tens}{S_{ij}}
\newcommand{\LS}{\bvec{L}_{ij}\cdot\bvec{S}_{ij}}
\newcommand{\TT}{\Tg_i\cdot\Tg_j}
\newcommand{\chet}{\ensuremath{\chi ET}}
\newcommand{\chpt}{\ensuremath{\chi PT}}
\newcommand{\chsy}{\ensuremath{\chi\mbox{symm}}}
\newcommand{\lchi}{\ensuremath{\Lambda_\chi}}
\newcommand{\lcon}{\ensuremath{\Lambda_{QCD}}}
\newcommand{\dcpsi}{\ensuremath{\bar{\psi}}}
\newcommand{\dc}[1]{\ensuremath{\overline{#1}}}
\newcommand{\llo}{\ensuremath{\mathcal{L}^{(0)}_{\chet}}}
\newcommand{\lchet}{\ensuremath{\mathcal{L}_{\chi}}}
\newcommand{\hchet}{\ensuremath{\mathcal{H}_{\chi}}}
\newcommand{\Dmu}{\ensuremath{\mathcal{D}_\mu}}
\newcommand{\Dsl}{\ensuremath{\slashed{\mathcal{D}}}}
\newcommand{\comm}[2]{\ensuremath{\left[#1,#2\right]}}
\newcommand{\acomm}[2]{\ensuremath{\left\{#1,#2\right\}}}
\newcommand{\third}{{\frac{1}{3}}}
\newcommand{\sthird}{{\scriptstyle\frac{1}{3}}}
\newcommand{\exv}[1]{\ensuremath{\bb\hat{#1}\kb}}
\newcommand{\ev}[1]{\ensuremath{\bb{#1}\kb}}
\newcommand{\pd}{\partial}
\newcommand{\pnpd}[2]{\frac{\partial{#1}}{\partial{#2}}}
\newcommand{\pppd}[1]{\frac{\partial{\hphantom{#1}}}{\partial{#1}}}
\newcommand{\plmu}{\partial_\mu}
\newcommand{\plnu}{\partial_\nu}
\newcommand{\pumu}{\partial^\mu}
\newcommand{\punu}{\partial^\nu}
\newcommand{\mcdf}{\delta^{(4)}(p_f-p_i-q)}
\newcommand{\ecdf}{\delta(E_f-E_i-\nu)}
\newcommand{\tr}{\mbox{Tr }}
\newcommand{\lxr}{\ensuremath{SU(2)_L\times SU(2)_R}}
\newcommand{\gV}[2]{\ensuremath{(\gamma^{-1})^{#1}_{\hphantom{#1}{#2}}}}
\newcommand{\gVd}[2]{\ensuremath{\gamma^{#1}_{\hphantom{#1}{#2}}}}
\newcommand{\LpV}[1]{\ensuremath{\Lambda^{#1}V}}
\newcommand{\hatH}{\ensuremath{\hat{H}}}
\newcommand{\Oop}{\ensuremath{\mathcal{O}}}

\title{Variational Monte Carlo study of pentaquark states in a correlated
quark model}


\author{Mark W. Paris}{
  address={Theory Group, Thomas Jefferson National Accelerator Facility,
  12000 Jefferson Avenue, MS 12H2, Newport News, Virginia, 23606}
}

\begin{abstract}
Accurate numerical solution of the five-body Schr\"{o}dinger
equation is effected via variational Monte Carlo in a correlated
quark model. The spectrum is assumed
to exhibit a narrow resonance with strangeness $S=+1$. A
fully antisymmetrized and pair-correlated five-quark
wave function is obtained for the assumed non-relativistic Hamiltonian
which has spin, isospin, and color dependent pair interactions and many-body
confining terms which are fixed by the non-exotic spectra.
Gauge field dynamics are modeled via flux tube exchange
factors.  The energy determined for the ground states with
$J^\pi=\shalf^-(\shalf^+)$ is 2.22 GeV (2.50 GeV).
A lower energy negative parity state is consistent with recent
lattice results.
\end{abstract}

\maketitle


A system of interacting, non-relativistic constituent quarks is the
most simple, realistic model of hadronic systems. Solving the many-body
Schr\"{o}dinger equation to determine wave functions within this simple 
model is still a formidable task owing to the strong flavor, spin, and 
color dependence of the quark-quark interaction and traditionally 
requires some array of approximate methods to solve it.
The controversial status of the recent evidence of a flavor exotic
five-quark state warrants a careful treatment of this strongly
interacting system. We have determined the wave function of five 
interacting constituent quarks in the flavor-exotic multiquark hadronic
sector is solved using the variational Monte Carlo (VMC) technique
\cite{Paris:2005sv}. This technique is known
to yield upper bounds on the ground state energy accurate to the level
of a few percent in light nuclei with the number of nucleons $A\le 6$
\cite{Pudliner:1995wk}.

Recent experimental evidence (for reviews see
\cite{Dzierba:2004db,Hicks:2005gp}) has revived interest 
in the multiquark flavor-exotic sector of the hadronic spectrum.
Various model calculations have been
reported (an overview is given in \cite{Oka:2004xh}) which study the 
existence of a strangeness $S=+1$ resonance, dubbed $\theta^+$, about 
100 MeV above threshold to nucleon-kaon decay with low mass of 1540(2) 
MeV and possibly extremely narrow width of $0.9(3)$ MeV
\cite{Eidelman:2004wy}. Lattice results are available, including Refs.
\cite{Mathur:2004jr,Lasscock:2005tt}.
High statistics photoproduction data of the reaction
$\gamma p \rightarrow \bar{K}^0 K^+ n$ from the CLAS 
collaboration \cite{diVita}
sets an upper limit on the yield of the $\theta^+$ relative 
to the $\Lambda^*(1520)$ yield at 0.2\%. No definitive structure in
the $nK$-invariant mass spectrum is observed at 1540 MeV in this
experiment. No experimental information is available on the spin or 
parity of the state in any experiment done to date.

We study the $\theta^+$ in the non-relativistic flux tube 
quark model with one-gluon exchange (OGE) and one-pion exchange (OPE) 
(between the light quarks). We take seriously the possibility that 
$\theta^+$ is a narrow resonance and calculate its mass as a stable
state with respect to strong interactions. The approach adopted here 
will be to work with a general, completely antisymmetric wave function 
and determine dynamically which flavor-spin-color-orbital
(TSCL) structures are favored in the constituent quark model (CQM).
The model Hamiltonian used in this work is
fixed by the single hadron spectrum \cite{Paris:2000phd} and the
six-quark (deuteron) properties \cite{Paris:2000bj}. It is given as a sum 
of kinetic energy, pair potentials for OGE and OPE, and confining terms.
The two-body operator potentials for OGE (OPE) interactions are 
determined in the non-relativistic reduction of the tree level 
amplitude for $qq$ or $q\bar{q}$ scattering. We use a many-body
confining interaction based on the ``flux tube'' model \cite{Isgur:1984bm}.
We consider flux tube topologies (there are three) for the $4q\bar{
q}$ states consistent with gauge invariance. Matrix elements of operators
which connect different flux tube topologies are modeled by a quark 
position dependent factor which models the dynamics of the gauge 
fields \cite{Paris:2000bj}.

The state vector for the five quarks, including two--body 
isospin, spin, color and spatial correlations is given by
\begin{equation}
\label{eqn:Psi5}
\ket{\Psi_5} = \mathcal{S}\prod_{i<j}\Fij\ket{\Phi_5}
\end{equation}
where the symmetrized product of two--body correlation
operators \Fij\ acts on an uncorrelated TSCL state, $\ket{\Phi_5}$.
The state $\ket{\Phi_5}$ and, therefore, $\ket{\Psi_5}$ is completely 
antisymmetric under exchange of the coordinates of the $u$ and $d$ 
quarks.

We considered a large set of states obtained by considering states
of total isospin $T=0$, $S=0,1$, and $L=0,1$ to yield states of negative
and positive parity.
The negative parity state $\ket{1;1^-}$ corresponds to the state of 
$4q$ with positive parity, $\pi_4=+1$ with zero orbital angular momentum.
The positive parity states, $\ket{n=1,\ldots,4;J_4^{+}}$ correspond to 
states of $4q$ with negative parity and include one unit of
orbital angular momentum. Here $J_4^{\pi}$ denotes the spin of the four 
light quarks, $J_4$ and the parity, $\pi$ of the pentaquark state.

\begin{table}[t]
\begin{tabular}{c|c@{\hspace{4mm}}c@{\hspace{4mm}}c@{\hspace{4mm}}c@{\hspace{4mm}}c@{\hspace{4mm}}c@{\hspace{4mm}}c}
\hline
$n;J_4^\pi$&$1;1^-$	&$1;1^+$&$2;1^+$&$3;0^+$&$3;1^+$&$4;0^+$&$4;1^+$\\
\hline
$M_{\theta^+}$&	2.22	&2.50	&2.57	&2.75	&2.81	&2.83	&2.88	\\
$\exv{T}$ &	1.68	&2.13	&2.02	&2.03	&2.00	&1.92	&1.90	\\
$\exv{V}$ &	0.92	&0.74	&0.93	&1.10	&1.19	&1.29	&1.36
\end{tabular}
\caption{\label{tab:Evmc} Variational energies of the states of
pentaquark states in GeV. The mass is reported as 
$M_{\theta^+}(n;J_4^\pi) =\exv{T} + \exv{V} - 385.5(5)$ MeV. 
Statistical errors for $M_{\theta^+}$ are $<5$ MeV.}
\end{table}

The evaluation of expectation values of the variational wave function
$\ket{\Psi_5}$ are effected via Monte Carlo integration of the 15 
dimensional space $\bvec{R}=(\rvec_1,\ldots,\rvec_5)$. Results for 
the variational energies of negative and positive parity states are 
shown in Table \ref{tab:Evmc}. The mass of the state is reported as
\begin{equation}
\label{eqn:mass}
M_{\theta^+}(n;J_4^{\pi}) = 4m_q + m_s + \exv{T} + \exv{V} - V_0(4q\bar{q})
\end{equation}
where $\exv{T}$ and $\exv{V}$ are the variational kinetic and potential
energies, respectively, for the states considered. The quark masses are
taken as $m_q=313$ MeV, $m_s=550$ MeV and $V_0$ is a constant fit to
the nucleon mass.  In reality, the
true positive parity ground state is some admixture of the six positive 
parity states. However, the lowest energy state dominates 
since the states $\ket{n;J_4^\pi}$ are orthonormal: applying the 
correlations as in Eq.(\ref{eqn:Psi5}) introduces off-diagonal elements 
which are quite small.  States with $n\ne 1$ contribute $<10\%$ to the 
true positive parity ground state.

Comparison of the CQM with lattice results at large pion mass is 
meaningful since the effects of chiral symmetry are negligible.
We note that our negative parity state $\ket{1;1^-}$ lies below the 
positive parity state $\ket{1;1^+}$, consistent with the lattice data
in Refs.\cite{Mathur:2004jr,Lasscock:2005tt}. This is true despite 
the fact that the positive parity state has much stronger attraction 
from OGE and OPE contributions, attributable to the
high degree of symmetry of the $TS$ state (see Table \ref{tab:Evmc}).
The $\exv{V}$ of the negative parity state is 180 MeV below the lowest
positive parity state. Exciting one unit of orbital angular momentum on 
the other hand raises the kinetic energy significantly, about 450 MeV.


We have shown that within the non-relativistic flux tube CQM with
OGE and OPE interactions the negative parity state is lower
than the positive parity state by $\sim 280$ MeV when the $\theta^+$
is assumed to be a narrow resonance. Though the higher symmetry of
the lowest lying positive parity state significantly decrease its 
potential energy, its unit excitation of orbital angular momentum 
raises the energy above that of the negative parity state. Quark
correlations are crucial in making these determinations and should
not be ignored.


\begin{theacknowledgments}
The author would like to thank Jozef Dudek, Bob Wiringa, and Ross Young
for helpful discussions. This work was supported by DOE contract 
DE-AC05-84ER40150 Modification No. M175, under which the Southeastern 
Universities Research Association (SURA) operates the Thomas Jefferson 
National Accelerator Facility.
\end{theacknowledgments}



\bibliographystyle{aipproc}   

\bibliography{paris05}

\IfFileExists{\jobname.bbl}{}
 {\typeout{}
  \typeout{******************************************}
  \typeout{** Please run "bibtex \jobname" to optain}
  \typeout{** the bibliography and then re-run LaTeX}
  \typeout{** twice to fix the references!}
  \typeout{******************************************}
  \typeout{}
 }

\end{document}